\documentclass{article}

\usepackage{arxiv}

\usepackage[utf8]{inputenc}
\usepackage[T1]{fontenc}
\usepackage{hyperref}
\usepackage{url}
\usepackage{booktabs}
\usepackage{amsfonts}
\usepackage{nicefrac}
\usepackage{microtype}
\usepackage{graphicx}
\graphicspath{{./figures/}}

\usepackage{amsmath}
\usepackage{pifont}
\usepackage{textcomp}
\usepackage{xcolor}
\usepackage{listings}
\usepackage{tikz}
\usepackage{multirow}
\usepackage{tabularx}
\usepackage{array}
\usepackage{makecell}
\usepackage{placeins}
\usepackage{float}
\usepackage{seqsplit}

\DeclareUnicodeCharacter{202F}{ }
\usetikzlibrary{shapes.geometric,arrows,fit,backgrounds,positioning}

\tikzstyle{process} = [rectangle, minimum width=2.7em, minimum height=2.5em, text centered, draw=black, fill=gray!10]
\tikzstyle{startend} = [ellipse, minimum width=2.7em, minimum height=1.5em, text centered, draw=red, fill=gray!10]
\tikzstyle{arrow} = [thick,->,>=stealth]

\lstdefinestyle{json}{
  basicstyle=\ttfamily\footnotesize,
  breaklines=true,
  frame=single,
  backgroundcolor=\color{gray!10},
  showstringspaces=false
}

\lstdefinestyle{cstyle}{
  language=C,
  basicstyle=\ttfamily\footnotesize,
  breaklines=true,
  frame=single,
  backgroundcolor=\color{gray!10},
  showstringspaces=false,
  keywordstyle=\color{blue},
  commentstyle=\color{green!50!black},
  stringstyle=\color{orange}
}

\providecommand{\Description}[1]{}

\newenvironment{acks}{\section*{Acknowledgments}}{}

\title{Library-Aware Doubles and Iterative Repair for Large Language Model-Generated Unit Tests in OpenSIL Firmware}

\author{
  Ma Toan Bach \\
  School of Software Design and Data Science\\
  Seneca Polytechnic\\
  Canada \\
  \href{mailto:matoanbach@gmail.com}{matoanbach@gmail.com} \\
  \And
  Yuchi Zheng \\
  School of Software Design and Data Science\\
  Seneca Polytechnic\\
  Canada \\
  \href{mailto:andyzheng.ca@gmail.com}{andyzheng.ca@gmail.com} \\
  \And
  Haingo Razafindranto \\
  School of Software Design and Data Science\\
  Seneca Polytechnic\\
  Canada \\
  \href{mailto:haingo.razafindranto@gmail.com}{haingo.razafindranto@gmail.com} \\
  \And
  Tanvir Alam \\
  School of Software Design and Data Science\\
  Seneca Polytechnic\\
  Canada \\
  \href{mailto:tanvir.alam@senecapolytechnic.ca}{tanvir.alam@senecapolytechnic.ca} \\
  \And
  Aric Leather \\
  Advanced Micro Devices\\
  Canada \\
  \href{mailto:Aric.Leather@amd.com}{Aric.Leather@amd.com} \\
  \And
  Ranveer Sandhu \\
  Advanced Micro Devices\\
  Canada \\
  \href{mailto:Ranveer.Sandhu@amd.com}{Ranveer.Sandhu@amd.com} \\
  \And
  Jitesh Arora \\
  Advanced Micro Devices\\
  Canada \\
  \href{mailto:Jitesh.Arora@amd.com}{Jitesh.Arora@amd.com} \\
}

\begin{document}
\maketitle

\begin{abstract}
Validating changes in low-level C firmware is expensive because unit tests (UTs) are fragile under strict build constraints, where missing headers, unresolved symbols, and dependency mismatches frequently prevent compilation and linking. This study introduces an automated UT authoring workflow for the Open-Source Silicon Initialization Library (openSIL) firmware codebase maintained by Advanced Micro Devices (AMD) that reduces manual effort through a large language model (LLM) guided multi-agent pipeline. The workflow combines automated generation of test scaffolds, library-aware creation or reuse of stubs, mocks, and fakes, and an iterative compile-dispatch repair loop driven by build logs and line-coverage feedback. We evaluate the approach using compilation success, repair iterations, dispatch success, and line coverage, with time, cost, and token usage as secondary measures. Across 76 functions under test, the workflow generated compilable UTs for 73 functions. In a configuration without line coverage guidance or retrieval augmentation, mean line coverage reached 73.9\%. On a 48-function subset evaluated under both configurations, mean line coverage reached 98.8\% with line-coverage guidance alone and reached 94.7\% when combined with vector-database retrieval. Results show that automated generation-and-repair pipelines can substantially improve UT creation efficiency and coverage for constrained firmware environments while reducing manual debugging effort.
\end{abstract}


\section{Introduction}
Unit tests (UTs) are a core mechanism for validating incremental software changes, and their effectiveness is commonly summarized using coverage metrics such as line and branch coverage \cite{Yang2024,Rahmani2020}. However, even when coverage is high, determining whether outputs are correct remains challenging because of the test oracle problem \cite{Barr2015}. These challenges are amplified in low-level firmware code, where tests must compile and link under strict build systems and where hardware-facing dependencies complicate isolation.

This challenge is especially relevant for the Open-Source Silicon Initialization Library (openSIL), an open-source firmware framework maintained by Advanced Micro Devices (AMD) for platform silicon initialization. Authoring UTs for openSIL requires tests to build under the Unified Extensible Firmware Interface (UEFI) Development Kit II (EDK II) constraints \cite{TianocoreEDK} while exercising firmware functions in isolation. In practice, tests often fail to build because of missing headers, unresolved symbols, incorrect build metadata (INF files), or mismatched package dependencies. Additionally, many openSIL functions call services in other modules, hardware-facing application programming interfaces (APIs), or function-pointer dispatch tables such as XFER tables and ip2ip APIs. Testing such functions typically requires test doubles, including stubs, mocks, or fakes, that both satisfy linker requirements and steer execution down the desired path. In C, this is typically done by adding hook points (seams) and redirecting calls at link time or through wrappers \cite{Feathers2004}. In addition, many functions under test (FUTs) rely on pointer-heavy dataflow, out-parameters, or memory-mapped register abstractions, increasing the likelihood of invalid preconditions and fragile doubles.

While large language models (LLMs) can draft readable UT scaffolds \cite{Yang2024,Schaefer2024}, an LLM alone is typically insufficient in this setting. The generated code must obey specific constraints such as templates, include rules, and symbol hygiene, and must be iteratively corrected using concrete build feedback. Prior studies on LLM-based UT generation have reported low ``compile-first'' rates and therefore relied on compile--execute--repair loops to obtain usable tests \cite{Pan2025,Yuan2024}. For firmware C UTs, these requirements are even stronger: rule enforcement to prevent duplicate production symbols, forbidden redefinitions, and template violations, and a tight feedback loop driven by compiler/linker logs are required before coverage improvements can be meaningfully pursued.

To address these challenges, this paper presents a workflow using multiple specialized LLM agents: some draft the initial UT files, others fix build issues using compiler/linker logs, and coverage reports from the line-coverage report generator (LCOV) are used to guide further improvements toward unexecuted lines \cite{LCOV}. The objective is to reduce manual intervention while achieving high build success, successful dispatch, and strong line coverage under firmware build constraints.

To guide the evaluation of this workflow, this study addresses the following research questions:
\begin{itemize}
  \item \textbf{RQ1:} How often does the workflow produce compilable and linkable UTs under EDK II?
  \item \textbf{RQ2:} How many repair iterations are typically needed to reach compile success and then progress to dispatch-ready, coverage-refinement iterations?
  \item \textbf{RQ3:} Under the same stopping rules, how does a workflow variant using line coverage analysis (LCA) only compare to one augmenting LCA with a vector database (VDB) in terms of line coverage and resource profile (runtime, cost, and token usage)?
\end{itemize}

This paper makes the following contributions:
\begin{itemize}
  \item A multi-agent UT authoring workflow for \texttt{openSIL} C that iteratively transitions UTs from initial drafts to compilable, dispatchable, and finally coverage-refined tests.
  \item A library-aware doubles strategy with explicit rule enforcement under EDK II constraints, reusing existing stub/mock/fake libraries when available and otherwise synthesizing minimal doubles without violating template or symbol-hygiene rules.
  \item An automated compile--dispatch repair loop that consumes compiler/linker logs and dispatcher outputs to apply targeted, minimal edits until tests reach maximum line coverage, aligning with prior compile--run--repair paradigms \cite{Pan2025,Yuan2024}.
  \item An empirical evaluation on 76 \texttt{openSIL} functions reporting compilation success rate, iterations-to-green, line coverage, and observed failure types, including comparisons of workflow variants with and without LCA feedback.
\end{itemize}

\section{Background}
Unit testing remains a cornerstone of software quality assurance, enabling early detection of defects and improving long-term maintainability. Its effectiveness is often quantified using metrics such as line coverage, branch coverage, and function coverage \cite{Rahmani2020,Staats2011}. While these metrics have been widely studied in high-level software systems, achieving high coverage in firmware or low-level C/C++ environments presents unique challenges. Firmware typically operates under strict build constraints (e.g., EDK II) \cite{TianocoreEDK}, involves pointer-intensive logic, and interacts directly with hardware registers, making isolation of testable units more difficult. In such settings, dependency isolation is typically achieved through C-level seams (e.g., link-time interposition or wrapper-based substitution) \cite{Feathers2004,GNUld}, and practical UT ecosystems commonly rely on dedicated C test frameworks (e.g., Ceedling/Unity/CMock) \cite{Ceedling}.

Automated UT generation has traditionally relied on symbolic execution \cite{Cadar2008}, mutation testing \cite{Jia2011}, and search-based approaches such as evolutionary test generation \cite{Fraser2011}. Symbolic execution offers path exploration guarantees but struggles with state-space explosion common in hardware-interfacing code \cite{Cadar2008}. Search-based methods are scalable but can produce brittle tests that break with minor code changes, and they often prioritize coverage rather than long-term maintainability \cite{Fraser2011}. As a result, most prior automation efforts have concentrated on high-level managed languages \cite{Pan2025,Deljouyi2025,Dakhel2024}, with only limited work targeting compiled-language firmware code \cite{Karanjai2024,QZhang2025}.

In recent years, large language models (LLMs) have emerged as an alternative for automated test generation, producing human-readable code and adapting flexibly to various programming paradigms \cite{Schaefer2024,Wang2024}. However, most prior research has employed non-reasoning models, which---while sometimes capable of chain-of-thought behavior---lack the explicit optimization for complex, multi-step problem solving seen in reasoning-tuned models \cite{Minaee2024}. In contrast, reasoning-enhanced models such as OpenAI's o3 \cite{OpenAI2025} incorporate chain-of-thought-style training signals to improve logical consistency and multi-step planning, making them strong candidates for multi-iteration code generation and repair \cite{Minaee2024}.

\subsection{Terminology}
The following terms are used throughout the paper:
\begin{itemize}
  \item \textit{Function under test (FUT):} the target C function for which a UT package is generated.
  \item \textit{Test double:} a replacement for a dependency during testing (for example, a stub, mock, or fake).
  \item \textit{Deep double:} a behavior-carrying replacement for cross-module calls made by the FUT.
  \item \textit{Shallow stub:} a minimal replacement for sibling-call fallout that exists mainly to satisfy compilation and linking.
  \item \textit{XFER table:} a function-pointer table used by openSIL modules for indirection and dispatch.
  \item \textit{Ip2Ip API:} an interface table containing function pointers returned at runtime and used for cross-IP calls.
  \item \textit{Chroma embedding:} conversion of source artifacts (for example, functions, existing tests, and test-double libraries) into vector representations stored in a Chroma database so semantically similar artifacts, as well as key-based or metadata-matched artifacts, can be retrieved during generation.
\end{itemize}

\subsection{Codebase Overview}
Open-Source Silicon Initialization Library (openSIL) is an open-source, modular firmware library primarily designed to initialize and configure AMD silicon platforms. It abstracts hardware-specific initialization logic, facilitating reuse across different firmware implementations, including coreboot and Unified Extensible Firmware Interface (UEFI)-based systems.

The codebase consists of distinct layers:
\begin{itemize}
\item \textit{Platform abstraction layer}: Provides standardized interfaces to abstract platform-specific functionalities, enhancing portability and code reuse.
\item \textit{Silicon Initialization Modules}: Includes functions that handle detailed initialization tasks for components like memory controllers, security processors, and chipset interfaces.
\item \textit{Configuration and Policy Management}: Encapsulates configurable parameters and policies influencing initialization logic.
\end{itemize}

In our workflow, a generated UT package for a single function consists of the following artifacts:
\begin{itemize}
\item \textit{UT header.} Holds most of the \texttt{openSIL} includes and shared declarations the test needs. When test doubles are needed, their declarations are typically placed here. This file is the primary place to manage which \texttt{openSIL} headers the test includes.
  \item \textit{UT source.} Contains the test cases and helper code. It includes the \textit{UT header} first; any additional harness headers are included only within template-approved include regions (e.g., inside designated double/helper blocks).
\item \textit{JavaScript Object Notation (JSON) file.} Describes the test inputs used for each test run (each iteration).
  \item \textit{Information file (INF).} Tells the EDK II build system what to compile and which packages/libraries the UT needs \cite{TianocoreEDK}.
\end{itemize}

{\small
\begin{figure}[tbp]
  \centering
  \begin{lstlisting}[style=cstyle]
typedef struct {
    DF_FABRIC_REGISTER_ACC_READ   DfFabricRegisterAccRead;
    DF_FABRIC_REGISTER_ACC_WRITE  DfFabricRegisterAccWrite;
    .
    .
    .
    DF_HAS_FCH                    DfHasFch;
    DF_HAS_SMU                    DfHasSmu;
} DF_COMMON_2_REV_XFER_BLOCK;

DF_COMMON_2_REV_XFER_BLOCK DfCmn2RevPhxXfer = {
    .DfFabricRegisterAccRead = DfXFabricRegisterAccRead,
    .DfFabricRegisterAccWrite = DfXFabricRegisterAccWrite,
    .
    .
    .
    .DfHasFch = PhxHasFch,
    .DfHasSmu = PhxHasSmu
};
  \end{lstlisting}
  \Description{C code listing that defines an XFER table struct (a function-pointer table) and initializes a global instance with pointers to platform-specific implementations.}
  \caption{Example \texttt{XFER} table structure.}
  \label{fig:xfer_struct}
\end{figure}
}

\begin{figure}[tbp]
  \centering
  \begin{lstlisting}[style=cstyle]
extern DF_COMMON_2_REV_XFER_BLOCK DfCmn2RevPhxXfer;
extern DF_IP2IP_API DfIp2IpApiPhx;

SIL_STATUS
InitializeApiDfXPhx (
  SIL_CONTEXT  *SilContext
  )
{
  SIL_STATUS Status;

  // Set Cmn2Rev table for DF
  Status = SilInitCommon2RevXferTable(SilContext, SilId_DfClass, &DfCmn2RevPhxXfer);
  if (Status != SilPass) {
    return Status;
  }

  // Set Ip2Ip API for DF
  return SilInitIp2IpApi(SilContext, SilId_DfClass, (void *) &DfIp2IpApiPhx);
}
  \end{lstlisting}
  \Description{C code listing of an initialization function that registers an XFER table and an Ip2Ip API table with a firmware context, returning early on failure.}
  \caption{Example use of an \texttt{XFER} table during initialization.}
  \label{fig:xfer_example}
\end{figure}

\begin{figure}[tbp]
  \centering
  \begin{lstlisting}[style=cstyle]
RAS_IP2IP_API *RasApi;
Status = SilGetIp2IpApi (SilId_RasClass, (void **)&RasApi);
if (Status != SilPass) {
  XPRF_TRACEPOINT (SIL_TRACE_ERROR, "RAS API not found!\n");
  return Status;
}
RasApi->ProgramCoreMcaIpIdInstanceId (RasCpuInfo);
  \end{lstlisting}
  \Description{C code listing that retrieves an Ip2Ip API table, checks the returned status, and calls a function through a function-pointer field of the retrieved table.}
  \caption{Retrieving an \texttt{Ip2IpApi} table and calling through a function pointer.}
  \label{fig:ip2ip_example}
\end{figure}

\subsection{Characteristics of Functions}
Functions in the \texttt{openSIL} codebase exhibit specific characteristics that influence the approach required to test them. The most relevant in our workflow relate to dependencies, output mechanisms, size, and linkage visibility. These characteristics include the following:

\textit{Functions that require doubles (deep vs.\ shallow)}: These functions may require deep doubles, which replace cross-module dependencies called by the FUT, or shallow stubs, which are minimal definitions for sibling-call fallout (functions in the same source file that are not under test) to satisfy compilation and linking without expanding behavior.

\textit{Functions that utilize an XFER table}: Certain functions rely on an XFER table, which is a function-pointer table used for indirection (see Figure~\ref{fig:xfer_struct} and Figure~\ref{fig:xfer_example}). To test these functions, \texttt{MockSilGetCommon2RevXferTableOnce}, a one-shot mock helper from the UT support library, configures the next call to \texttt{SilGetCommon2RevXferTable} to return a test-controlled XFER table and status code.

\textit{Functions that utilize Ip2Ip functionality}: Some functions use an Ip2Ip structure containing function pointers (see Figure~\ref{fig:ip2ip_example}). To test these functions, \texttt{MockSilGetIp2IpApiOnce}, a corresponding one-shot mock helper, configures the next call to \texttt{SilGetIp2IpApi} to return a test-controlled Ip2Ip table and status code.

\textit{Functions that return a value}: Some functions return a value via a return statement. In such cases, the return type is identified from the signature, and correctness is evaluated using assertions on the return value (noting that oracle strength remains a general challenge) \cite{Barr2015}.

\textit{Functions that return a value through a pointer argument}: Other functions return values through pointer arguments. These cases are validated by asserting on modified output parameters.

\textit{Static functions}: Static functions have internal linkage and cannot be called from outside their source file. Because our UTs are built as separate UT packages, static-only targets require an explicit seam (e.g., a non-static wrapper or a harness that exposes the behavior); otherwise, they are excluded as FUTs.

\textit{Small Functions}: typically fewer than 20 lines of code.

\textit{Medium Functions}: range from 20 to 70 lines of code.

\textit{Large Functions}: contain more than 70 lines of code.

The above size categories use lines of code (LOC) as a pragmatic proxy for code size, a common software measurement practice \cite{Fenton2014}.

\section{Related Work}
Prior work related to this paper falls into four areas: (i) using LLMs to generate UTs, (ii) fixing generated tests by looping over compiler or test-run errors, (iii) using retrieval-augmented generation to bring in relevant code context during generation, and (iv) C/systems testing practices, especially test doubles and link-time build constraints. We structure the discussion using five criteria: target language/build constraints, compilation/execution feedback loop, retrieval of project context (including reusable doubles), explicit build/link error handling and coverage-driven refinement.

\subsection{Criterion 1: Target Language and Build Constraints (C/Systems)}
Most LLM-based UT generation work focuses on Java, JavaScript/TypeScript, or Python, where builds usually do not involve strict linking rules and dependency wiring is relatively simple \cite{Yang2024,Schaefer2024,Yuan2024}. However, some work targets compiled languages (e.g., C++ UTs) \cite{Karanjai2024,YZhang2025}, but firmware build systems add extra constraints: tests must follow EDK II rules for library classes, allowed include placement, and avoiding symbol conflicts \cite{TianocoreEDK}. In C, dependencies are often replaced using test hook points (seams), for example by changing which symbol the linker binds to or by routing calls through wrappers \cite{Feathers2004,GNUld}. Test harnesses are also commonly built with C frameworks such as Ceedling/Unity/CMock \cite{Ceedling}. This paper focuses on generating openSIL C firmware UTs while respecting these firmware-specific build constraints.

\subsection{Criterion 2: Compilation and Execution Feedback Loops}
A common result in prior studies is that tests generated in a single LLM pass often fail to compile initially \cite{Yang2024}. Because of this, several systems use a loop: generate tests, attempt to compile and run them, then fix the code using the errors as guidance \cite{Pan2025,Yuan2024}. A multi-language LLM-based UT generation and repair workflow named ASTER follows this approach and uses static-analysis outputs plus compile/run results to drive repeated fixes \cite{Pan2025}. ChatTester, a Java-focused generate--compile--run--repair approach, also automates repeated rounds based on build and test outcomes \cite{Yuan2024}. More broadly, this matches the idea of treating tool outputs (compiler errors, test failures, logs) as the most reliable signal to constrain and correct generated code \cite{Wang2024}. We use the same idea, but in a firmware setting: compilation means an EDK II build, execution means running the UT dispatcher, and repairs are limited to safe edits that follow the desired UT templates and avoid symbol conflicts.

\subsection{Criterion 3: Retrieval of Existing Doubles and Code Context}
Many recent systems improve LLM-generated code by retrieving project-specific context from a codebase or document store, instead of relying only on the model's built-in knowledge \cite{Lewis2020,RepoCoder2023,CodeRAG2025}. In UT generation, this often means pulling in example tests or other project hints to make the output match real code \cite{Pan2025,Lemieux2023}.

In our workflow, retrieval is used as a practical ``reuse step'' before writing new code. It retrieves (1) related functions from the codebase, (2) existing UTs for those related functions, and (3) existing helper libraries for test doubles (stubs/mocks/fakes) for the FUT's dependencies. This helps in two main ways. First, it reduces build and link failures by avoiding duplicate definitions and by filling in missing dependencies with helpers that already work in this environment. Second, it reduces guesswork by reusing examples that already build, so the workflow can copy the right \texttt{\#include} lines and build settings instead of trial-and-error. Details of the three retrieval queries and collections are described in Section~\ref{subsubsec:retrieval}.

\subsection{Criterion 4: Handling Build/Link Errors Explicitly}
In C, UTs often fail to build for very concrete reasons: a missing \texttt{\#include}, a missing function declaration, or the linker finding duplicate symbols (redefinition error) or a missing symbol (unresolved symbol) \cite{GNUld,Feathers2004}. Prior work in program repair treats compiler and test output as the ``signal'' that guides fixes \cite{Weimer2009,Monperrus2018}. Similarly, ASTER and ChatTester read build errors and iterate until the generated tests compile \cite{Pan2025,Yuan2024}.

Our approach follows the same idea: compiler and linker messages decide what must be fixed next. For firmware, we add two practical safeguards: (1) a retrieved ``do-not-redefine'' list to avoid generating symbols that must not be reimplemented, and (2) a fixed set of \texttt{\#include} lines provided by retrieved libraries \cite{TianocoreEDK}.

\subsection{Criterion 5: Coverage-Driven Refinement}
Coverage numbers indicate which lines of code were executed, but they do not prove that a test checks the ``right'' behavior unless the assertions are strong (the oracle problem) \cite{Staats2011,Barr2015}. Nevertheless, coverage can be a practical guide for improving tests: some systems use an LLM in combination with coverage feedback to get past points where coverage stops increasing (e.g., CodaMosa) \cite{Lemieux2023}, and others use analysis results to decide the next action (e.g., ASTER) \cite{Pan2025}.

In our workflow, after a test runs, we read LCOV's per-line results \cite{LCOV} and build a simple map that marks which lines were hit and which were missed. The next edits focus on the missed lines---for example by adding or adjusting input iterations or setup---until a fixed stopping rule is reached (coverage threshold or iteration limit).

\section{Materials and Methods}

\subsection{Dataset}
The codebase used in this study is the publicly available \texttt{openSIL} project hosted on GitHub: \url{https://github.com/openSIL/openSIL/tree/phoenix_poc} (accessed on 29 April 2026). The \texttt{phoenix\_poc} branch was selected as the evaluation target, and source C files from this branch were used as targets for UT generation in all experiments.

\subsection{Human Workflow}
To provide a reference point, we first describe the manual process used by developers to create UTs in \texttt{openSIL}. First, developers select a function within a source file and create a dedicated UT folder. Next, they add the standard files including an information file (INF), an iteration JSON file, and the UT source and header files. Then, they register the new test by adding the path to its UT INF file to a centralized registry for UT libraries (AmdOpenSilUtPkg.dsc.inc).

Then, developers attempt to build the test. If the build fails, developers are required to fix the reported problems one by one. Common issues when creating UTs are missing headers and missing or duplicate symbols. To fix them, developers need to determine the FUT's dependencies, then either link an existing library or create a local stub/mock/fake, update the \texttt{\#include}s and the INF settings, and rebuild. They repeat this process until the test compiles and links.

Once the test builds, the developer writes the actual test assertions in the UT source file and adds the first set of inputs (iteration and description) in the UT JSON file. They then rebuild and run the test to obtain a dispatcher report (pass/fail status) and a coverage report (e.g., LCOV) \cite{LCOV}. If the line coverage result is low, developers add or refine test inputs and cases to hit more branches. This process repeats in a cycle: edit the UT code, rebuild, rerun, and review coverage again, until the target coverage percentage is reached.

\subsection{System Overview}
We propose a retrieval-augmented iterative pipeline for automated UT generation. The workflow consists of an initial UT generation phase followed by an automated repair loop driven by compiler, linker, and coverage feedback.

The system relies on two core components. First, a VDB implemented using Chroma stores and retrieves existing functions (knowledge base), existing UTs (ground truth), and reusable test doubles (libraries). Second, LCA is used after successful execution to produce a per-line hit/miss map from LCOV output, which guides subsequent refinement steps.

The workflow proceeds as follows:
\begin{itemize}
  \item \textbf{Retrieval phase.} Related functions (knowledge base), their UT code (ground truth), and reusable doubles (libraries) are extracted from a VDB.
  \item \textbf{Drafting phase.} Produce the first version of the UT header, UT source, INF, and iteration JSON.
  \item \textbf{Assembly phase.} Merge the draft into the template-approved layout and add only the include or helper code that the template allows.
  \item \textbf{Build and execution phase.} Compile the test under EDK II, then run the dispatcher when the build succeeds.
  \item \textbf{Repair phase.} Apply small edits using compiler/linker errors, dispatcher output, and coverage feedback until a set limit on attempts or when coverage passes a chosen threshold.
\end{itemize}

The system maintains a structured loop that continues until either a coverage threshold is reached or a maximum number of iterations is exceeded.

\subsection{Execution Environment}
The workflow is implemented in \textbf{Python} (version $\geq$ 3.8). \textbf{LangGraph StateGraph} handles the workflow orchestration as a set of steps (nodes) with iterative loops, while the LLM interactions are managed through \textbf{LangChain ChatOpenAI}. Settings (such as paths and model choices) are loaded from \textbf{dotenv} and a \textbf{toml} configuration file.

\subsection{VDB Construction}

\label{subsubsec:retrieval}

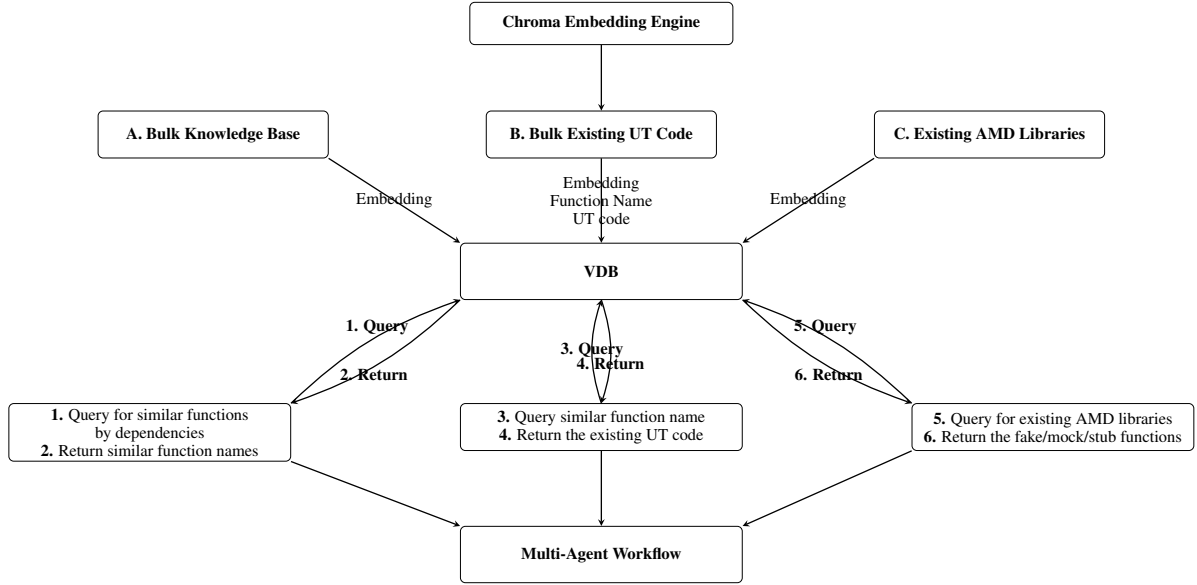
\begin{figure}[tbp]
\centering
\scriptsize
\resizebox{0.95\textwidth}{!}{%
\begin{tikzpicture}[
  font=\normalsize,
  node distance=8mm and 10mm,
  box/.style={draw, rounded corners=3pt, align=center, minimum height=9mm, inner sep=3pt},
  sbox/.style={box, minimum width=4.9cm, minimum height=10mm},
  db/.style={box, minimum width=6.0cm, minimum height=12mm},
  wbox/.style={box, minimum width=6.0cm, minimum height=12mm},
  qbox/.style={box, minimum width=6.0cm, minimum height=10mm},
  lab/.style={align=center, inner sep=1pt},
  arrow/.style={->,>=stealth, thick},
  darrow/.style={->,>=stealth, thick, double, double distance=0.9pt}
]

\node (eng) [box, minimum width=5.6cm] {\textbf{Chroma Embedding Engine}};

\node (frameA) [sbox, below left=14mm and 30mm of eng] {\textbf{A. Bulk Knowledge Base}};
\node (frameB) [sbox, below=14mm of eng] {\textbf{B. Bulk Existing UT Code}};
\node (frameC) [sbox, below right=14mm and 30mm of eng] {\textbf{C. Existing AMD Libraries}};

\draw[arrow] (eng.south) -- (frameB.north);

\node (vdb) [db, below=18mm of frameB] {\textbf{VDB}};

\draw[arrow] (frameA.south east) -- node[midway,lab]{Embedding} (vdb.north west);
\draw[arrow] (frameB.south) -- node[midway,lab]{Embedding\\ Function Name\\ UT code} (vdb.north);
\draw[arrow] (frameC.south west) -- node[midway,lab]{Embedding} (vdb.north east);

\node (qleft)  [qbox, below left=22mm and 36mm of vdb] {
  \textbf{1.} Query for similar functions\\
  by dependencies\\
  \textbf{2.} Return similar function names
};

\node (qmid)   [qbox, below=22mm of vdb] {
  \textbf{3.} Query similar function name\\
  \textbf{4.} Return the existing UT code
};

\node (qright) [qbox, below right=22mm and 36mm of vdb] {
  \textbf{5.} Query for existing AMD libraries\\
  \textbf{6.} Return the fake/mock/stub functions
};

\node (ma) [wbox, below=16mm of qmid] {\textbf{Multi-Agent Workflow}};

\draw[arrow] (qleft.north east) to[bend left=12]
  node[pos=0.55, above, lab]{\textbf{1. Query}} (vdb.south west);
\draw[arrow] (vdb.south west) to[bend left=12]
  node[pos=0.55, below, lab]{\textbf{2. Return}} (qleft.north east);

\draw[arrow] (qmid.north) to[bend left=18]
  node[pos=0.45, lab, yshift=4pt]{\textbf{3. Query}} (vdb.south);

\draw[arrow] (vdb.south) to[bend left=18]
  node[pos=0.55, lab, yshift=-4pt]{\textbf{4. Return}} (qmid.north);

\draw[arrow] (qright.north west) to[bend right=12]
  node[pos=0.55, above, lab]{\textbf{5. Query}} (vdb.south east);
\draw[arrow] (vdb.south east) to[bend right=12]
  node[pos=0.55, below, lab]{\textbf{6. Return}} (qright.north west);

\draw[arrow] (qleft.south east) -- (ma.north west);
\draw[arrow] (qmid.south) -- (ma.north);
\draw[arrow] (qright.south west) -- (ma.north east);

\end{tikzpicture}%
}
\Description{Diagram of the retrieval subsystem: a Chroma embedding engine ingests three sources (a KnowledgeBase JSON, existing UTs, and existing test-double libraries) into a VDB; three query boxes retrieve similar functions, matching UTs, and reusable double libraries, feeding a multi-agent workflow.}
\caption{VDB integration for retrieval: Embeddings are generated for (i) knowledge base entries, (ii) existing UTs, and (iii) AMD test-double libraries. During generation, the workflow issues three retrieval queries: dependency-based similarity for related functions (1--2), UT matching by function name (3--4), and test-double library suggestions for unresolved dependencies (5--6).}
\label{fig:vectordb_retrieval}
\end{figure}

\noindent
The workflow uses a Chroma-based VDB to store and retrieve project context during generation. As shown in Figure~\ref{fig:vectordb_retrieval}, it maintains three Chroma collections and issues one targeted query per collection:
\begin{itemize}
  \item \textbf{Knowledge base:} dependency-similarity search to suggest functions similar to the FUT (steps 1--2).
  \item \textbf{Ground truth:} retrieve existing UT code for those related functions as examples (steps 3--4).
  \item \textbf{Library doubles:} suggest reusable stubs/mocks/fakes and their known-good \texttt{\#include} directives (steps 5--6).
\end{itemize}

During execution, the system performs three retrieval queries: (i) dependency-based retrieval of related functions, (ii) matching of existing UTs for similar functions, and (iii) retrieval of reusable test doubles for unresolved dependencies. Retrieved artifacts are injected into later generation and repair stages to ensure consistency with existing project structure and reduce invalid builds.

\subsection{Workflow Orchestration}
\begin{figure}[tbp]
\centering
{\scriptsize
\resizebox{0.86\textwidth}{!}{%
\begin{tikzpicture}[node distance=2.2em and 1.6em, every node/.style={font=\scriptsize}]
\node (START) [startend] {Start};

\node (LOAD) [process, below=of START] {\parbox[t][][t]{2.6cm}{\centering \texttt{load}\\\scriptsize validate + init state}};
\node (RDDD) [process, below=of LOAD] {\parbox[t][][t]{3.6cm}{\centering \texttt{retrieve\_deep\_doubles\_node}\\\scriptsize VDB $\rightarrow$ retrieved libs + rules}};
\node (WUT)  [process, below=of RDDD] {\parbox[t][][t]{3.0cm}{\centering \texttt{write\_unittest\_node}\\\scriptsize draft \texttt{ut\_c}/\texttt{ut\_h} + \texttt{ut\_library\_classes}}};
\node (WSS)  [process, below=of WUT] {\parbox[t][][t]{3.6cm}{\centering \texttt{write\_shallow\_stub\_node}\\\scriptsize sibling stubs + \texttt{sibling\_library\_stubs\_map}}};
\node (ASM)  [process, below=of WSS] {\parbox[t][][t]{2.8cm}{\centering \texttt{assemble\_node}\\\scriptsize merge templates + enforce ban list}};
\node (RES)  [process, below=of ASM] {\parbox[t][][t]{2.8cm}{\centering \texttt{resolve\_node}\\\scriptsize compile-fix; then dispatch-driven refinement}};
\node (SET)  [process, below=of RES] {\parbox[t][][t]{3cm}{\centering \texttt{setup\_node}\\\scriptsize extract iterations; create UT folder; merge base libs}};
\node (COM)  [process, below=of SET] {\parbox[t][][t]{2.4cm}{\centering \texttt{compile\_node}}};
\node (DIS)  [process, below=of COM] {\parbox[t][][t]{2.4cm}{\centering \texttt{dispatch\_node}}};

\node (BP2)  [diamond, draw=black, fill=gray!10, text width=3.4cm, aspect=2,
             below=of DIS, yshift=-0.8em] {\parbox[t][][t]{3.6cm}{\centering \texttt{break\_point\_2}\\\scriptsize continue? (interactive/non-interactive)\\\scriptsize coverage threshold + iteration limit}};
\node (OUT)  [process, right=7.8em of BP2] {\parbox[t][][t]{2.1cm}{\centering \texttt{output}}};

\draw [arrow] (START) -- (LOAD);
\draw [arrow] (LOAD) -- (RDDD);
\draw [arrow] (RDDD) -- (WUT);
\draw [arrow] (WUT) -- (WSS);
\draw [arrow] (WSS) -- (ASM);
\draw [arrow] (ASM) -- (RES);
\draw [arrow] (RES) -- (SET);
\draw [arrow] (SET) -- (COM);
\draw [arrow] (COM) -- (DIS);
\draw [arrow] (DIS) -- (BP2);
\draw [arrow] (BP2) -- node[anchor=south]{No} (OUT);

\draw [arrow] (BP2.west) |- node[anchor=east,pos=0.25]{Yes} (RES.west);

\end{tikzpicture}%
}
}
\Description{Workflow state graph with 11 steps in a vertical pipeline and a loop: load, retrieve deep doubles, draft UT, generate shallow stubs, assemble, resolve, setup, compile, dispatch, break-point decision, and output, with a loop from the decision back to resolve.}
\caption{11-stage workflow. The loop is controlled by \texttt{break\_point\_2}, which supports both interactive and non-interactive continuation based on coverage thresholds and iteration limits.}
\label{fig:workflow_graph}
\end{figure}
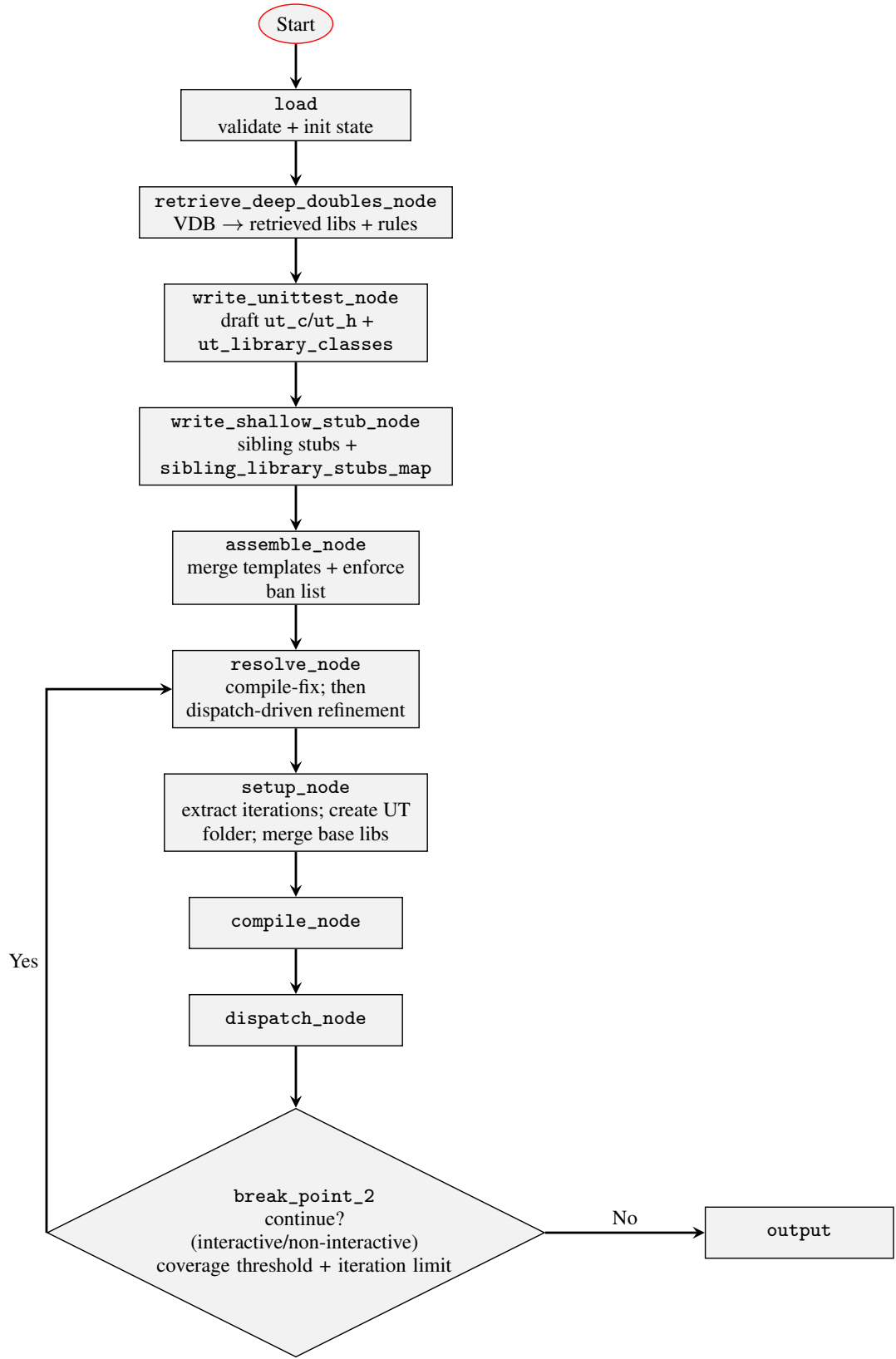

The automated workflow is implemented as a state graph consisting of 11 steps (Figure~\ref{fig:workflow_graph}). It can operate in either interactive or non-interactive mode. In interactive mode, a user may review intermediate outputs and provide feedback; in non-interactive mode, execution is fully automated and governed by predefined stopping criteria.

Some steps perform normal automation work (reading/writing files, running the build, running the test). Other steps call an LLM to write or update the UT files. Each node performs a specific role in the pipeline:

{\small
\begin{itemize}
  \item \textbf{\texttt{load}} --- Initializes workflow state and validates inputs.
  \item \textbf{\texttt{retrieve\_deep\_doubles\_node}} --- Retrieves existing UT helper libraries (stubs/mocks/fakes) and associated usage constraints, along with the needed \texttt{\#include} lines in the UT header of dependencies used by the FUT.
  \item \textbf{\texttt{write\_unittest\_node}} --- Retrieves example UT code for functions with similar dependencies (via knowledge base and ground-truth lookup through the VDB) and uses them as an ``example tests'' block while drafting \texttt{UT source} and \texttt{UT header}, along with the needed INF library dependencies.
  \item \textbf{\texttt{write\_shallow\_stub\_node}} --- Generates minimal stubs or retrieves existing UT libraries for sibling functions in the same source file as the FUT, and uses any library context provided for them.
  \item \textbf{\texttt{assemble\_node}} --- Combines all code into the final file layout and applies the project rules (e.g., do not redefine forbidden symbols).
  \item \textbf{\texttt{resolve\_node}} --- Fixes problems found by the build or test run (compile errors, link errors, test run issues, or coverage gaps) using small edits.
  \item \textbf{\texttt{setup\_node}} --- Creates the UT folder/files and registers the test with the build system.
  \item \textbf{\texttt{compile\_node}} --- Builds the UT with EDK II and saves the compiler/linker output \cite{TianocoreEDK}.
  \item \textbf{\texttt{dispatch\_node}} --- Runs the UT, saves the run logs, and generates coverage using LCOV \cite{LCOV}, which may be used as the input for LCA in later refinement rounds.
  \item \textbf{\texttt{break\_point\_2}} --- Decides whether to stop or perform another fix/refine round based on conditions such as coverage thresholds and iteration limits. In non-interactive mode, this step is automatically handled based on the target coverage (default 90\% or above) or a set iteration limit, both of which are configurable. In interactive mode, this node allows users to provide feedback to the LLM, which will be used to guide the next fix/refinement step.
  \item \textbf{\texttt{output}} --- Finalizes execution.
\end{itemize}
}

This design follows the same general idea as prior ``generate, run tools, fix, repeat'' workflows \cite{Pan2025,Yuan2024}. The key difference for \texttt{openSIL} firmware is that fixes are limited to safe places in the templates, and the workflow avoids redefining symbols or adding headers in ways that break EDK II builds.

\subsection{Design Rationale}
\subsubsection*{Why Split Deep Doubles Versus Shallow Stubs}
\textbf{Pain point:} A firmware UT can fail to link for two different reasons: (1) the FUT calls code in other modules, and (2) other functions in the same source file also invoke extra calls. If we treat every missing function the same way, we often ``fix'' the build by adding code that clashes with existing libraries (duplicate symbols) or breaks the required templates.

\textbf{Rationale:} We handle these two cases separately. For cross-module calls, we create \emph{deep doubles} that behave similarly to the real dependency enough to let the test reach the intended code path. For extra calls introduced by sibling functions, we create \emph{shallow stubs} that are minimal and only exist to make the test build and link safely.

\subsubsection*{Why Retrieve Existing Doubles First}
\textbf{Pain point:} In a large UT codebase, helper code for dependencies often already exists. If we write a new stub/mock/fake for a function that already has one, the build may fail because the linker encounters two versions of the same symbol. Even when it still builds, different tests can result in the use of slightly different helpers for the same dependency, which makes the test suite harder to keep consistent.

\textbf{Rationale:} The workflow looks for an existing helper first and uses it when available. This reduces duplicate-symbol build errors and reuses helper code. We describe how this lookup works in section~\ref{subsubsec:retrieval}, specifically in the Test-double library collection.

\subsubsection*{Why an Iterative Repair Loop Is Required}
\textbf{Pain point:} In \texttt{openSIL} C UTs, it is normal for the first draft not to build or run. The compiler and linker often report errors related to missing headers, missing libraries in the INF, or unresolved symbols. Even after the test builds, it may still fail at runtime due to errors in syntax, setup steps, or input conditions.

\textbf{Rationale:} The workflow follows the same loop a developer would use: write an initial test, attempt to build it, review the errors, address the reported issues, and repeat. After it builds, the workflow runs the test, reads the dispatcher output, and fixes runtime issues in the same manner. Each iteration uses build and execution logs to determine the next incremental modification, rather than attempting to resolve all issues in a single step \cite{Pan2025,Yuan2024}.

\subsubsection*{Why a Coverage Gate in \texttt{break\_point\_2}, and a Line Coverage Threshold in Non-Interactive Mode}
\textbf{Pain point:} In interactive mode, a user can choose whether to continue additional repair/refinement rounds; in non-interactive mode, the workflow runs autonomously and stops based on configured thresholds (e.g., a coverage target and an iteration limit). If we keep iterating only to increase coverage, the workflow may stall, spending many rounds on the last few lines in non-interactive mode. Those lines are often behind rare error paths, hardware-dependent checks, or guards that are difficult to trigger in a UT environment.

\textbf{Rationale:} \texttt{break\_point\_2} is a clear stopping rule that prevents the workflow from running indefinitely. In non-interactive mode, the workflow stops once it reaches a default target of at least 90\% line coverage (this value is configurable). The goal is to achieve sufficient coverage efficiently, while still acknowledging that coverage is only a proxy and does not guarantee correctness by itself \cite{Staats2011,Barr2015}.

\subsection{Prompting and Guardrails}

\subsubsection*{Prompts}
Each LLM stage in the workflow uses a modular prompt template with the same general organization, while the content of each section is specialized to the task at hand.

The prompt template is organized into five recurring components:
\begin{itemize}
  \item \textbf{Identity}: This section states the role the model should assume for the current stage, such as drafting UTs, generating shallow stubs, retrieving or validating deep doubles, or repairing code after build and dispatch feedback. The identity section frames the stage so that the model knows what kind of artifact it is expected to produce.
  \item \textbf{Inputs}: This section lists the values passed to the model for that stage. Typical inputs include the FUT source file, function name, dependency lists, retrieved library guidance, build logs, dispatcher output, coverage summaries, and any stage-specific metadata. The input set changes by stage so that the model receives only the context it needs for the current task.
  \item \textbf{Steps}: This section provides the stage-by-stage procedure the model should follow. For example, a shallow-stub stage may first filter sibling dependencies, then extract signatures, then generate minimal stub bodies, and finally validate the result before output. Other stages use the same idea but with different ordered actions, such as drafting, repair, or coverage-guided refinement. The purpose of this section is to make the execution order explicit and to prevent the model from skipping directly to generation before handling the required intermediate checks.
  \item \textbf{Validation}: This section tells the model how to check its own output before returning it. Validation commonly includes signature matching, dependency filtering, conformance with library guidance, and a final consistency check against the rules supplied for that stage.
  \item \textbf{Final Output}: This section specifies the required output from the LLM. Depending on the stage, the output may be a set of stubs, a revised UT file fragment, a repair patch, or a structured metadata object.
\end{itemize}

To keep generated UTs buildable in \texttt{openSIL}, the workflow enforces some constraints.

\subsubsection*{Symbol and Redefinition Control}
The firmware UTs can break if a stub/mock is written for a dependency that is already provided by an existing UT library used in the UT code. To prevent this, the workflow retrieves a ``do-not-define'' list created by the \texttt{retrieve\_deep\_doubles\_node} node and reviews it during file assembly and UT fixes. If a symbol is on the list, the workflow avoids generating a new definition for it.

\subsubsection*{Usage of Known-Good \texttt{\#include} Lines (Include Directives)}
The retrieved libraries output from the VDB provide the exact \texttt{\#include} lines that need to be included in the UT header file. This ensures consistency between library dependencies and UT compilation requirements in the EDK II environment.

\subsubsection*{Template-Constrained Generation}
The workflow relies on two fixed templates: one for the UT source file and one for the header file to keep generation and edits safe. These template files are consumed by the agents to produce the UT artifacts. The templates themselves are not compiled or executed as standalone UTs. The templates enforce:
\begin{itemize}
  \item \textbf{Where edits are allowed}: The templates mark specific \texttt{TODO} blocks where the workflow is allowed to add code (e.g., deep doubles, shallow stubs, and test-case logic). All edits are confined to these blocks.
  \item \textbf{Controlled include regions in the C file}: The generated UT source includes the UT header first; any additional harness headers are introduced only inside template-approved include regions (e.g., within designated helper/double blocks), rather than as unrestricted top-level includes.
  \item \textbf{A central include surface in the header}: The test header template includes a central section for all necessary includes. This is where openSIL project headers, unit test framework headers, and any required helper headers are collected to keep the test consistent and maintainable.
  \item \textbf{Separation of responsibilities}: The template keeps \emph{deep doubles} separate from \emph{shallow stubs}. It also forbids unsafe shortcuts such as redefining production symbols or stubbing static functions.
\end{itemize}

\subsubsection*{Fixed LLM Output Schema}
Each LLM step must return its output in a fixed format, so the workflow can reliably read and apply it. In our implementation, we use output schemas for:
\begin{itemize}
  \item Unit-test drafting: produces the main UT files (\texttt{ut\_c}, \texttt{ut\_h}) and key INF fields.
  \item Shallow stubs: produces stubs for sibling functions.
  \item Deep-doubles retrieval: returns which existing libraries were found and chosen, along with library usage rules that the workflow must follow.
  \item Repair/refinement edits: returns small, specific edits to apply when fixing build/test failures or improving coverage.
  \item Iterations: creates or edits the iteration JSON file.
\end{itemize}

\subsection{Coverage-Guided Iteration}
After the UT builds successfully, the workflow runs it and collects a line-by-line coverage report using LCOV \cite{LCOV}. The workflow also combines results across iterations into one merged view (the LCA map) that shows which lines were hit and which were missed.

If some test cases fail, or some lines are still not executed, the workflow does a small, targeted update in \texttt{resolve\_node}. Typical fixes include:
\begin{itemize}
  \item adding a new iteration to exercise a missing branch,
  \item adjusting inputs for an existing iteration, or
  \item setting up doubles so the function can reach the intended path.
\end{itemize}

This replaces the manual developer workflow of inspecting coverage reports and deciding on follow-up modifications into a repeatable, tool-driven step. It is similar in \emph{idea} to coverage-guided fuzzing \cite{Manes2021}---using coverage feedback to decide the next action---but here the workflow applies that feedback by editing structured UT artifacts (iterations and doubles) rather than mutating raw inputs.

Figure~\ref{fig:lcov_analysis} shows an example of the LCOV-derived hit/miss line mapping used by this step.

{\small
\begin{figure}[tbp]
  \centering
  \begin{lstlisting}[style=cstyle,
    breaklines=true,
    breakatwhitespace=false,
    columns=fullflexible,
    keepspaces=true,
    breakindent=0pt,
    postbreak=\mbox{\textcolor{gray}{$\hookrightarrow$}\space},
    basicstyle=\ttfamily\scriptsize]
59:: void
60::  IntToString (
61::    char       *String,
62::    uint8_t    *Integer,
63::    uint8_t    SizeInByte
64::  )
65:1: {
66::   uint8_t Index;
67::
68:1:   for (Index = 0; Index < SizeInByte; Index++) {
69:1:     *(String + Index * 2) =
69:1:       ( (*(Integer + Index) >> 4) & 0x0F );
70:1:     *(String + Index * 2 + 1) =
70:1:       ( *(Integer + Index) & 0x0F );
71:1:   }
72:1:   for (Index = 0; Index < (SizeInByte * 2); Index++) {
73:1:     if (*(String + Index) >= 0x0A) {
74:0:       *(String + Index) += 0x37;
75:0:     } else {
76:1:       *(String + Index) += 0x30;
77::     }
78:1:   }
79:1:   *(String + SizeInByte * 2) = 0x0;
80:1: }
  \end{lstlisting}
  \Description{Excerpt of a C function annotated with LCOV hit counts (1 for executed, 0 for not executed), illustrating how per-line coverage is mapped back to source lines for analysis.}
  \caption{C source with per-line hit/miss annotations derived from LCOV, where 1 = hit and 0 = miss.}
  \label{fig:lcov_analysis}
\end{figure}
}

\section{Experimental Setup}
\label{sec:experimental-setup}
We evaluated the workflow on AMD's openSIL firmware codebase. The study uses 76 FUTs. They were randomly sampled from the set of functions that can be executed under the UT framework and then grouped for reporting by size (Small/Medium/Large by LOC) and dependency patterns (including XFER and Ip2Ip). Static-only targets were generally excluded unless an explicit seam/exposure mechanism was available.

All experiments ran on a Windows machine with the EDK II build environment installed \cite{TianocoreEDK}. For each FUT, we ran the pipeline end-to-end to:
(1) generate the UT files (\texttt{.c}, \texttt{.h}, \texttt{.inf}, and iteration \texttt{.json}),
(2) build and link the test under EDK II,
(3) run the UT (dispatch) when the build succeeded, and
(4) compute coverage using LCOV \cite{LCOV}. Although build and dispatch run in the Windows-hosted EDK II environment, LCOV processing is executed through a POSIX-compatible environment (e.g., WSL) because the LCOV/gcov toolchain and path conventions are POSIX-oriented. The current workflow is integrated with a Windows-based EDK II UT setup; full support for native Linux environments was not part of the evaluated configuration in this study.

\textbf{Run-count.} The number of repeated runs for each FUT depends on the complexity of the function, but the default value used was 10 repetitions of the workflow. While Small and Medium functions generally achieved high line coverage within that number of runs, Large functions usually required more iterations.

\textbf{When the loop stops.} The workflow stops when any of the following happens:
(i) the non-interactive coverage target is reached (90\% or above by default),
(ii) coverage stops improving under \texttt{break\_point\_2}, or
(iii) the workflow reaches its set iteration limit.

\subsection{LLM Configuration}

The workflow was instantiated with three commercially available LLMs selected to balance capability and computational cost. Model selection was guided by public code-editing benchmarks \cite{Aider}, internal pilot runs, and the differing requirements of each workflow stage. Alternative models considered included other GPT-family variants and Gemini-family models; however, the selected configuration provided a practical balance of performance and efficiency for the tasks in this workflow.

\begin{itemize}
  \item \textbf{GPT-4.1-mini} is used in the shallow-stub step (\texttt{write\_shallow\_stub\_node}) to generate sibling stubs in batches, and in \texttt{setup\_node} to extract iterations from drafted UT files. These steps are mostly repetitive and template-driven.
  \item \textbf{o4-mini} is used in the drafting and assembly steps (\texttt{write\_unittest\_node}, \allowbreak  \texttt{assemble\_node}) to draft the main UT files and combine them into a working set of artifacts. This step needs consistent code across multiple files, while maintaining computational efficiency.
  \item \textbf{o3} is used in the repair/refinement step (\texttt{resolve\_node}) to fix problems found during build and test runs. It applies small edits based on compiler/linker errors and on dispatch/coverage logs, where careful reasoning helps avoid breaking the build.
\end{itemize}

\subsection{Evaluation Configurations}
We evaluate two workflow configurations under the same stopping rules (\allowbreak  \texttt{break\_point\_2}) and the same iteration/coverage thresholds, along with a direct LLM-only baseline for comparison:
\begin{itemize}
  \item \textbf{Direct LLM-only baseline:} prompt-based UT generation without structured repair or retrieval.
  \item \textbf{LCA-only:} LCA enabled; VDB retrieval disabled.
  \item \textbf{LCA+VDB:} LCA enabled; VDB retrieval enabled, so retrieved deep doubles and library usage rules are injected into drafting, assembly, and compile-fix stages.
\end{itemize}

\section{Results}
\label{sec:results}
This section presents the experimental results in alignment with the research questions, beginning with buildability under EDK II (RQ1), then examining how many repair iterations are needed to reach build and dispatch-ready refinement (RQ2), and finally comparing the LCA-only and LCA+VDB configurations under the same stopping rules (RQ3). To support this evaluation, we report results across four configurations: a direct LLM-only baseline, a no LCA/no VDB variant, an LCA-only variant, and an LCA+VDB variant. This comparative design allows us to isolate the effect of coverage guidance and retrieval-backed context while tracking build success, iteration counts, and coverage/resource trade-offs.

Coverage is only available when dispatch completes and LCOV artifacts are produced; some compilable UTs did not yield coverage within the configured dispatch/repair limits, which is why coverage sample sizes are smaller than the compile-success count.

\subsection{RQ1: Buildability of Generated Unit Tests}
Out of 76 FUTs, the workflow produced UTs that compile and link under EDK II for 73 cases (96.1\%). The remaining 3 FUTs did not achieve a successful build within the default repair limit of 10 iterations, typically due to unresolved deep dependencies or strict build constraints that could not be safely satisfied.

Table~\ref{tab:results-summary} breaks down outcomes by category for (i) a direct LLM configuration, (ii) a configuration without LCA or VDB retrieval, and (iii) the 48-FUT subset evaluated under both LCA-only and LCA+VDB.

Across this 48-FUT subset, mean line coverage is 98.8\% for LCA-only and 94.7\% for LCA+VDB (arithmetic mean of per-FUT percentages).

To contextualize these results, we include a direct LLM-only baseline computed on a representative subset without structured repair or retrieval. As shown in Table~\ref{tab:results-summary}, this baseline achieves substantially lower compile success across all categories, highlighting the importance of iterative repair and dependency-aware generation.

\begin{table}[t]
\centering
\small
\caption{Outcomes by FUT category. Direct LLM-only results are computed on a representative subset of FUTs and correspond to generation without structured repair or retrieval. LCA-only and LCA+VDB are computed on the 48-FUT subset evaluated in both configurations. Success counts the FUTs with compilable (compile and link) UTs. Avg Cov is the arithmetic mean of per-FUT LCOV line coverage percentages (unweighted by LOC) and is reported only for FUTs that successfully produce coverage outputs. Median Iters is the median number of resolve-loop iterations (including compile-fix and dispatch/coverage refinement); for the direct LLM-only baseline, iterations represent prompt attempts rather than structured repair-loop iterations.}
\label{tab:results-summary}
\begin{tabular}{>{\raggedright\arraybackslash}p{5cm} lccc}
\toprule
Configuration & Category & Success & Avg Cov (\%) & Median Iters \\
\midrule
\multirow{4}{*}{Baseline (Direct LLM-only)} & Small      & 15 &  55.2 & 5 \\
                                     & Medium     & 15 &  44.7 & 7 \\
                                     & Large      &  2 &  21.5 & 10 \\
                                     & XFER/Ip2Ip &  8 &  12.3 & 8 \\
\midrule
\multirow{4}{*}{No LCA/VDB} & Small      & 34 &  78.9 & 2 \\
                                     & Medium     & 31 &  70.7 & 2 \\
                                     & Large      &  8 &  71.3 & 4 \\
                                     & XFER/Ip2Ip  &  3 &  57.8 & 2 \\
\midrule
\multirow{4}{*}{LCA-only}            & Small      & 10 & 100.0 & 2 \\
                                     & Medium     & 10 & 100.0 & 2 \\
                                     & Large      &  8 &  94.3 & 6 \\
                                     & XFER/Ip2Ip  & 20 &  99.4 & 4.5 \\
\midrule
\multirow{4}{*}{LCA+VDB}             & Small      & 10 & 100.0 & 2 \\
                                     & Medium     & 10 &  99.6 & 3 \\
                                     & Large      &  8 &  79.9 & 9.5 \\
                                     & XFER/Ip2Ip  & 20 &  95.5 & 4 \\
\bottomrule
\end{tabular}
\end{table}

\subsection{RQ2: Repair Iterations and Failure Patterns}
In most cases, the workflow reaches compile success after a small number of iterations, and then performs additional iterations only when dispatch does not complete or when coverage remains below the stopping threshold. Table~\ref{tab:results-summary} shows that median iteration counts are generally low across configurations. Under the no LCA/no VDB configuration, median iterations are 2 for Small, Medium, and XFER/Ip2Ip categories, increasing to 4 for Large builds. With LCA-only, medians remain at 2 for Small and Medium, rise to 6 for Large, and reach 4.5 for XFER/Ip2Ip. Under LCA+VDB, median iterations are again 2 for Small, 3 for Medium, 9.5 for Large, and 4 for XFER/Ip2Ip. Overall, Small and Medium builds typically converge within two to three iterations, while Large builds require more retries, particularly when VDB is enabled. This suggests that most repair effort is concentrated in more complex builds, whereas simpler builds stabilize quickly. For the direct LLM-only baseline, iteration counts reflect prompt attempts rather than structured repair-loop iterations, which explains the higher median values observed in Table~\ref{tab:results-summary}.

\subsubsection*{Failure Patterns}
Error categories were extracted from compiler/linker diagnostics and dispatcher logs. The most common build issues are missing headers or prototypes, unresolved symbols at link time, duplicate symbol definitions, and signature mismatches. When a test builds but dispatch fails to complete, the usual causes are missing initialization steps or unmet runtime preconditions in the test setup.

\subsection{RQ3: Effect of LCA and VDB on Coverage and Resource Usage}
\begin{figure}[tbp]
  \centering
  \includegraphics[width=0.4\textwidth]{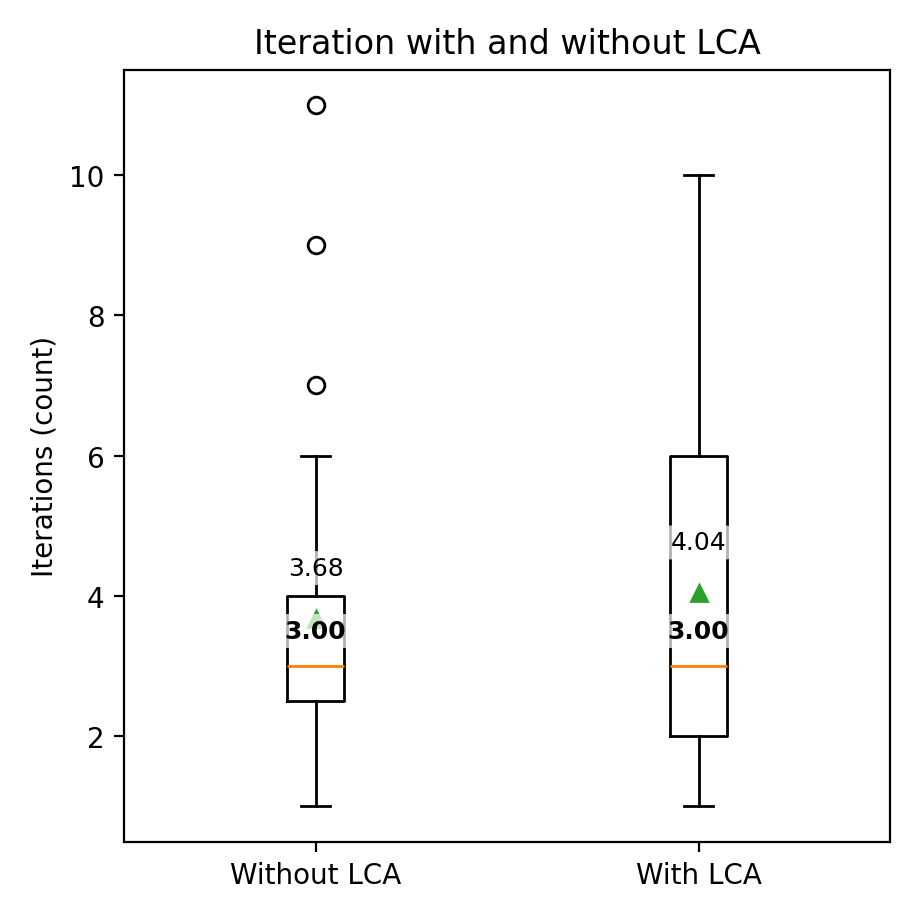}
  \Description{Boxplot-style figure comparing repair iterations, runtime, and cost between configurations with and without line-coverage analysis (LCA).}
  \caption{Iterations, time, and cost with and without LCA.}
  \label{fig:box_iterations}
\end{figure}

\begin{figure}[tbp]
  \centering
  \includegraphics[width=0.94\textwidth]{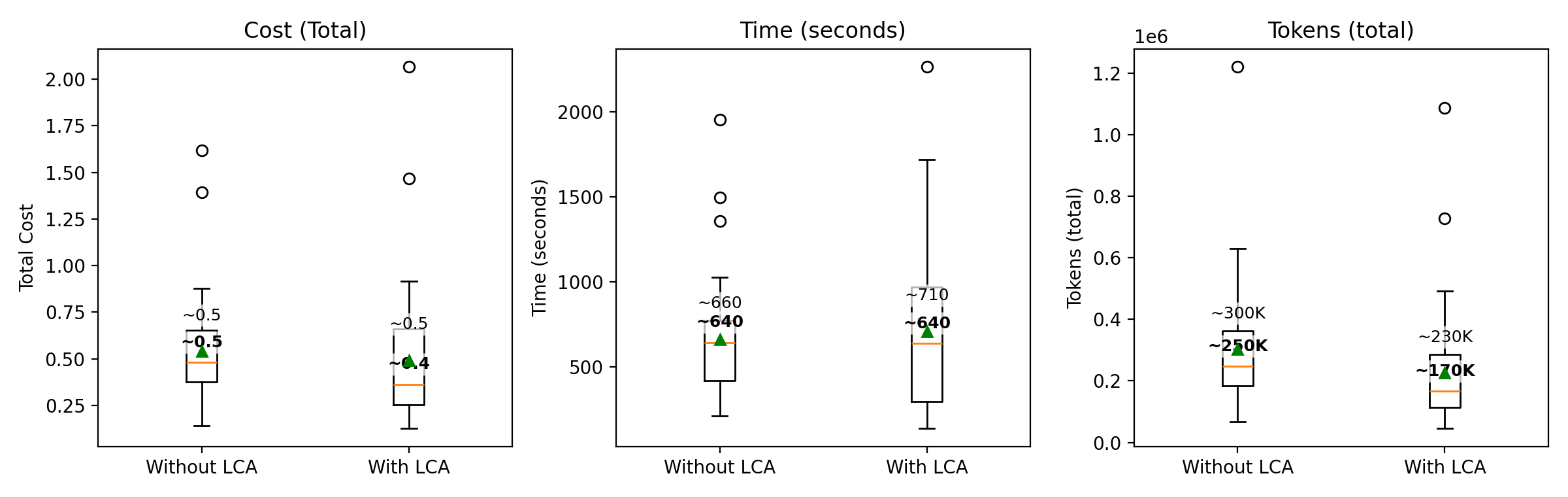}
  \Description{Boxplot-style summary figure comparing distributions of runtime, cost, and token usage between configurations with and without line-coverage analysis (LCA).}
  \caption{Distribution of time, cost, and tokens for configurations with and without LCA.}
  \label{fig:box_overview}
\end{figure}

Line coverage is computed from LCOV outputs \cite{LCOV}. In the no LCA/no VDB configuration, mean line coverage is 73.9\% over 76 compile-success FUTs. On the 48-FUT subset evaluated under both configurations, LCA-only reached 98.8\% mean line coverage and LCA+VDB reached 94.7\% mean line coverage (Table~\ref{tab:results-summary}). The key difference is that LCA points the repair step toward specific uncovered lines, which reduces trial-and-error when choosing inputs and priming doubles. Distribution summaries for iterations and for time/cost/tokens are shown in Figure~\ref{fig:box_iterations} and Figure~\ref{fig:box_overview}, respectively.

\subsection{VDB-Backed Library Integration Alters Time and Token Usage}
\begin{figure}[tbp]
  \centering
  \includegraphics[width=0.94\textwidth]{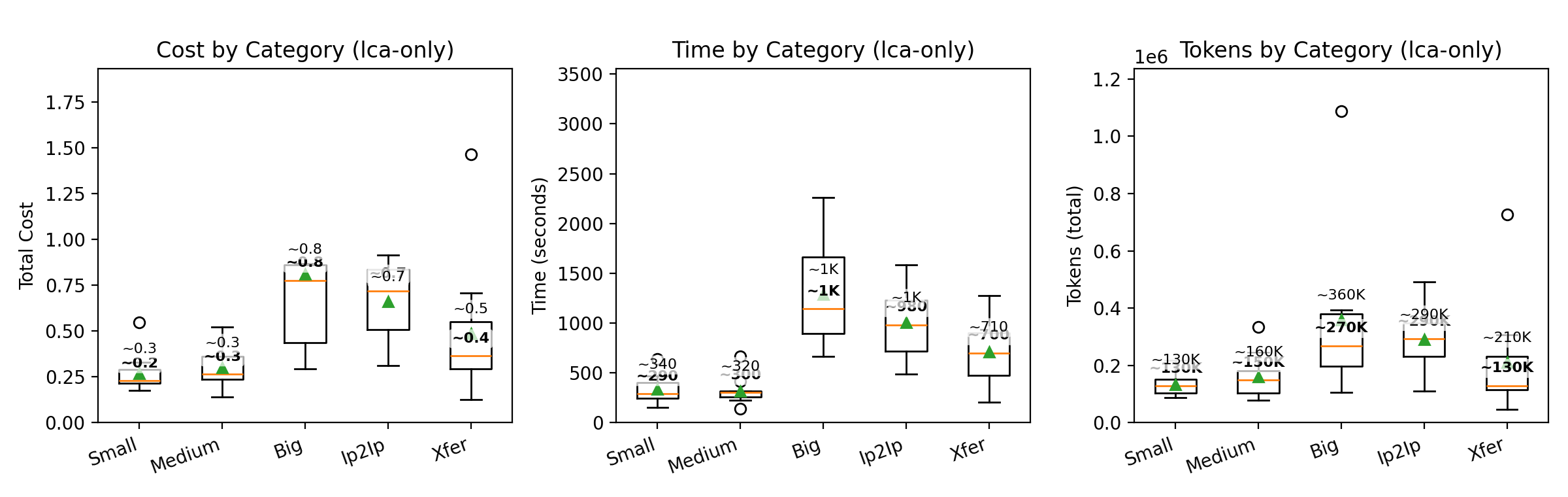}
  \Description{Bar/box-style comparison of time, cost, and total tokens by function category for the LCA-only configuration (LCA enabled without VDB retrieval).}
  \caption{Time (s), cost, and total tokens by function category for the \textbf{LCA-only} configuration (LCA enabled; no VDB retrieval).}
  \label{fig:lca-only}
\end{figure}

\begin{figure}[tbp]
  \centering
  \includegraphics[width=0.94\textwidth]{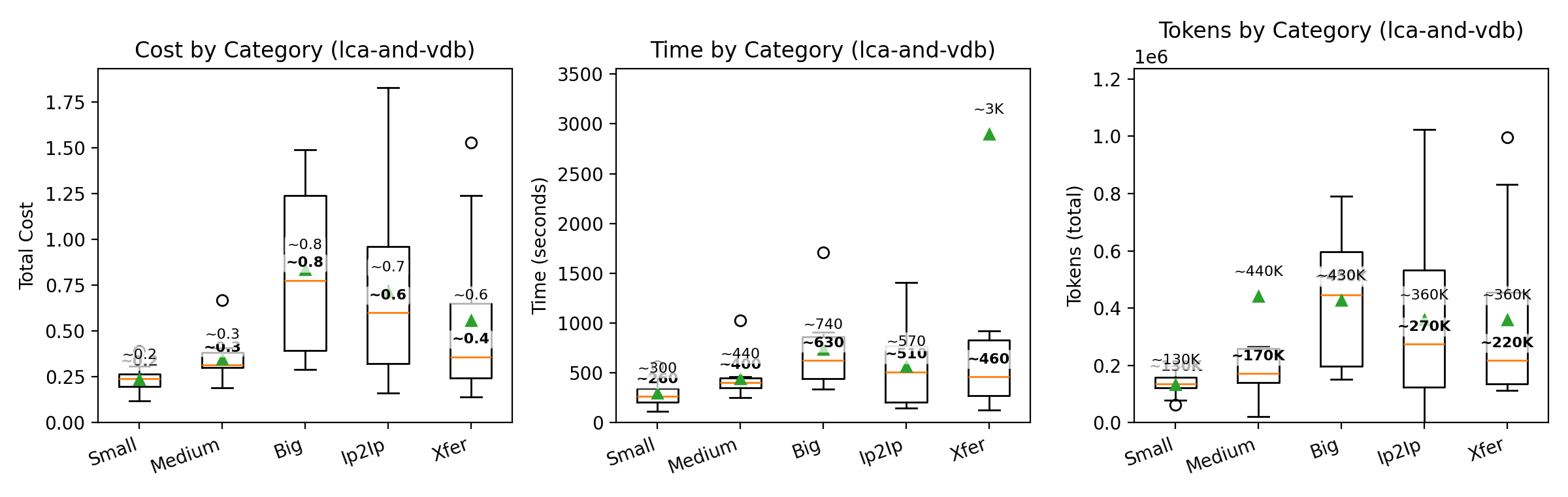}
  \Description{Bar/box-style comparison of time, cost, and total tokens by function category for the LCA+VDB configuration (LCA enabled with VDB retrieval).}
  \caption{Time (s), cost, and total tokens by function category for the \textbf{LCA+VDB} configuration (LCA enabled; VDB retrieval enabled).}
  \label{fig:lca-and-vdb}
\end{figure}

Figure~\ref{fig:lca-only} and Figure~\ref{fig:lca-and-vdb} compare results by function category for the LCA-only and LCA+VDB configurations, respectively. In the LCA+VDB plot, total runtime is much lower for more complex categories (for example, Large and Ip2Ip). At the same time, total tokens are higher and vary more in several categories (including Medium, Large, and XFER). The cost remains relatively steady across categories, with smaller changes than runtime and tokens.

When VDB retrieval is enabled, the workflow fetches two retrieval outputs (\allowbreak \texttt{retrieved\_deep\_doubles} and \texttt{library\_rules}) and incorporates them into multiple LLM prompts: drafting (\texttt{write\_unittest\_node}), assembly (\texttt{assemble\_node}), and build-fix (\texttt{resolve\_node}). These retrieved doubles and library rules provide lists of dependencies that should not be re-implemented, along with required \texttt{\#include} directives for the UT header file. As a result, prompt length and token usage may increase.

It cannot be claimed that VDB retrieval strictly reduces iteration counts. Rather, our results suggest that retrieval alters the runtime--token trade-off by introducing additional context and constraints, which can reduce wall-clock time for more complex categories while increasing total token usage.

\subsection{Human vs.\ Automated Workflow}
In informal developer observations, manual UT authoring involves repeated scaffold + compile-fix + coverage-fix cycles per function, where effort tends to increase with function size and dependency complexity. The automated workflow aims to reduce this manual iteration by systematically consuming build/dispatch/coverage feedback and applying targeted edits. A quantitative productivity comparison (e.g., UTs per day) is left as future work unless developer time-on-task is measured under controlled conditions.

\subsection{Comparison Across Function Categories}
The LCA-only configuration compares results by function type. Tests for Small and Medium functions are usually faster, cheaper, and more predictable. Large functions typically take more iterations and cost more. XFER/Ip2Ip cases are the hardest and vary the most, because the test often needs extra setup for pointer tables and more doubles.

In our runs, LCA helps most for Large and XFER/Ip2Ip functions by pointing the repair step to the lines that are still uncovered. For Small and Medium functions, execution is already fast and stable, so LCA has a smaller impact.

\section{Discussion}
\label{sec:discussion}
We applied the workflow to a real firmware codebase (AMD openSIL) and evaluated it end-to-end on 76 functions under test (FUTs). The discussion below summarizes what mattered most in practice.

\begin{itemize}
  \item \textbf{Buildability comes first.} While LLM-based approaches focus on generating test logic, our results show that many failures arise from build conventions and dependency wiring rather than from test intent. A UT is only useful if it compiles, links, and runs inside the existing toolchain; many failures are unrelated to test intent and come from build conventions and wiring. This contrasts with direct LLM-only generation, which lacks structured repair and often fails to produce buildable tests under the same constraints.
  \item \textbf{Library reuse reduces linker risk but can cost iterations.} Reusing stub/mock/fake libraries helps reduce linker errors and improves integration reliability, but integration effort can trade off against coverage under a fixed iteration budget. This indicates that effective reuse strategies are essential for scaling test generation in large codebases but must be balanced against iteration efficiency.
  \item \textbf{Coverage-guided feedback enables targeted refinement.} LCA transforms coverage gaps into concrete repair signals, reducing trial-and-error in test generation. This explains the significant coverage improvements observed with LCA-enabled configurations and suggests that feedback-driven refinement is critical for achieving high-quality tests in complex environments. In contrast, direct LLM-only approaches lack such feedback signals, leading to lower coverage and less targeted test refinement.
  \item \textbf{Most blockers are wiring, not algorithms.} Missing headers/prototypes, unresolved symbols, duplicate definitions, and signature mismatches dominate failures; dispatch failures often reflect missing setup/preconditions. This suggests that one of the main challenges in firmware UT generation lies in correctly wiring dependencies.
  \item \textbf{Additional context changes the runtime--token trade-off.} Introducing retrieval-based context using a VDB can reduce runtime for complex cases by guiding generation but increases token usage and variability. This suggests a trade-off between efficiency and context richness, which should be considered when designing LLM-based testing workflows.
\end{itemize}

\subsection{Generalizability Beyond openSIL}
The proposed workflow is not specific to openSIL but reflects a general pattern for LLM-assisted UT generation in complex C-based systems. This also suggests that approaches relying solely on direct LLM generation may not generalize well to environments with strict build and dependency constraints. In particular, the combination of context retrieval, iterative build--repair loops, and coverage-guided refinement addresses common challenges across firmware and systems-level codebases, such as dependency resolution, strict build constraints, and incomplete execution coverage.

The overall workflow design can be applied to other projects with similar characteristics, including large-scale C/C++ systems that rely on structured build pipelines and modular dependencies.

However, practical adoption depends on project-specific infrastructure, including build scripts, template constraints, available test doubles, and coverage/dispatch tooling. As a result, while the workflow structure generalizes, components such as prompts, retrieval rules, and library configurations may require adaptation to the target environment.

\subsection{Threats to Validity}
\label{sec:threats-validity}
\subsubsection*{Construct Validity}
Compile success, dispatch success, and LCOV line coverage are practical proxies for buildability and exercised code, but they do not prove semantic correctness or strong oracles.

\subsubsection*{External Validity}
The data used in this study are limited to AMD openSIL under EDK II constraints, so the results should not be assumed to generalize directly to other firmware stacks, languages, or build systems.

\subsubsection*{Ablation Validity}
This study does not explicitly isolate how the quality or availability of pre-existing UTs and test doubles affects outcomes. However, the quality of the generated UTs inherently depends on the correctness and completeness of the existing test doubles and UTs. Incorrect or incomplete doubles may lead to test failures or aborts, while flawed existing UTs, used as ground truth, could propagate errors into the generated tests. As such, conclusions about retrieval benefits should be interpreted as configuration-level observations rather than causal claims.

\subsubsection*{Conclusion Validity}
The evaluation uses a fixed set of 76 FUTs and a smaller subset for some comparisons. The reported conclusions therefore reflect the studied codebase and configuration rather than a broad claim about all UT generation settings.

\section{Conclusions}
This paper presented a multi-agent workflow for generating and improving UTs for a real-world firmware system (AMD openSIL) using build, dispatch, and coverage feedback. Compared to direct LLM-only generation, which often fails to satisfy build and dependency constraints, our workflow explicitly addresses practical challenges such as buildability, dependency resolution, and execution constraints in complex codebases.

Our evaluation demonstrates that the proposed approach can produce compilable and linkable UTs for 73 out of 76 functions while achieving high line coverage. These results indicate that integrating iterative repair and coverage-guided refinement is essential to making LLM-based test generation viable in real-world environments.

Moreover, our findings suggest that the primary challenges in firmware unit testing lie not in generating test logic but in ensuring correct integration with existing build systems.

At the same time, several limitations remain. Complex functions with deep dependencies and pointer-heavy patterns can still hinder performance. Addressing these limitations will require better modeling of cross-module dependencies and stronger analysis of pointer-heavy code paths.

Future work could extend the evaluation to additional firmware codebases and build environments to assess robustness beyond the openSIL setting. It could also improve the handling of complex functions by incorporating richer dependency analysis, particularly for cross-module interactions and pointer-heavy code paths. In addition, future studies could examine the impact of retrieval quality, such as the completeness and correctness of existing UTs and test doubles, on generation outcomes. They could also consider larger and more diverse sets of functions.

\section*{Author Contributions}
Conceptualization, J.A.; methodology, Y.Z., M.T.B. and H.R.; software, M.T.B., Y.Z. and H.R.; validation, M.T.B., Y.Z., H.R., T.A., A.L., R.S. and J.A.; formal analysis, M.T.B., Y.Z. and H.R.; investigation, M.T.B., Y.Z. and H.R.; resources, A.L., R.S. and J.A.; data curation, M.T.B., Y.Z. and H.R.; writing---original draft preparation, M.T.B.; writing---review and editing, H.R., Y.Z. and T.A.; visualization, M.T.B. and Y.Z.; supervision, T.A. and J.A.; project administration, T.A. and J.A.; funding acquisition, T.A. and J.A. All authors have read and agreed to the published version of the manuscript.

\section*{Funding}
This research was funded by the Natural Sciences and Engineering Research Council of Canada (NSERC) Mobilize grant and an NSERC Applied Research and Development grant (grant numbers CCMOB-2021-00190 and CCARD-2024-00774), with additional financial support from AMD. The authors would like to express their sincere gratitude to AMD managers and engineers for their valuable support throughout the development of this research.

\section*{Data Availability Statement}
The openSIL repository (\url{https://github.com/openSIL/openSIL/tree/phoenix_poc}) (accessed on 29 April 2026) was used as the codebase for unit test generation in this study and is distributed under the MIT License. A pseudocode representation of the workflow pipeline developed in this study is provided as Supplementary Material to illustrate the overall system architecture. Additional code used in this work cannot be made publicly available due to legal and contractual restrictions, as it is subject to AMD intellectual property rights and confidentiality agreements.

\begin{acks}
During the preparation of this manuscript, generative AI was used to assist with proofreading and manuscript preparation. The authors have reviewed all content and take full responsibility for the accuracy and integrity of this publication.
\end{acks}

\section*{Conflicts of Interest}
Authors Aric Leather, Ranveer Sandhu and Jitesh Arora are employed by AMD. The remaining authors declare no commercial or financial relationships that could be construed as potential conflicts of interest.

\bibliographystyle{unsrt}
\bibliography{references}

\end{document}